\documentclass{article}

\usepackage{siunitx}
\usepackage{subfigure}
\usepackage[section]{placeins}
\usepackage{xcolor}
\usepackage{adjustbox}
\usepackage{graphics}
\usepackage{float}
\usepackage{epsfig}
\usepackage{caption}
\usepackage{slashed}
\usepackage{color}
\usepackage{soul}
\usepackage{changes}
\usepackage{bbm}
\usepackage{pgfplots}
\usepgfplotslibrary{groupplots}
\usepgfplotslibrary{fillbetween}
\usepgfplotslibrary{colormaps}
\usepackage{tikzscale}
\usepackage{tikz}
\pgfplotsset{compat=1.15}
\usetikzlibrary{calc}
\usetikzlibrary{spy}
\usetikzlibrary{fadings} 
\usepgfplotslibrary{external}   
\usepackage{amsmath}
\usepackage{authblk}

\tikzfading[name=fade right,
  left color=transparent!0, right color=transparent!100]
\tikzfading[name=fade left,
  right color=transparent!0, left color=transparent!100]

\pgfplotsset{
    colormap={plainwhite}{
        rgb255=(255,255,255)
        rgb255=(255,255,255)
    },
    colormap={gray}{
        rgb255=(255,255,255)
        rgb255=(0,0,0)
    },
}

\pgfplotsset{
    every non boxed x axis/.style={} 
}

\pgfkeys{
  /pgf/number format/set thousands separator =
}

\title{Sample thickness measurements by phase-sensitive terahertz upconversion detection}
\date{}

\author[1,2,*]{Tobias Pfeiffer}
\author[1]{Jens Klier}
\author[1,2]{Georg von Freymann}
\author[1]{Daniel Molter}
\affil[1]{Fraunhofer-Institute for Industrial Mathematics ITWM, Fraunhofer-Platz 1, 67663 Kaiserslautern, GERMANY}
\affil[2]{Department of Physics and Research Center OPTIMAS, Technische Universit\"at Kaiserslautern (TUK), 67663 Kaiserslautern, GERMANY}
\affil[*]{tobias.pfeiffer@itwm.fraunhofer.de} %% email address is required 

\begin{document}

% Note that article type is not required for Express journals (OE, BOE, OME and OSAC)

\maketitle


\vspace{\baselineskip}


\begin{abstract}
Nonlinear frequency conversion provides an elegant method to detect photons in a spectral range which differs from the pump wavelength, making it highly attractive for photons with inherently low energy. Aside from the intensity of the light, represented by the number of photons, their phase provides important information and enables a plethora of applications. We present a phase-sensitive measurement method in the terahertz spectral range by only detecting visible light. Using the optical interference of frequency-converted photons and leftover pump photons of the involved ultrashort pulses, fast determination of layer-thicknesses is demonstrated. %Phase-sensitive detection of terahertz radiation is crucial for precise time-of-flight measurements due to the highly increased resolution compared to Intensity-only detection. In this work we present our accomplishments in phase-sensitive terahertz measurements by detecting visible light only, improving upon an established terahertz upconversion detection setup. 
The new method enables phase-resolved detection of terahertz pulses using standard sCMOS equipment while achieving sample measurement times of less than one second with a precision error of less than 0.6\%.
\end{abstract}

%%%%%%%%%%%%%%%%%%%%%%%%%%  body  %%%%%%%%%%%%%%%%%%%%%%%%%%
\section{Introduction}

The quantum-optical concept of measuring in extreme spectral ranges by using correlated photons in nonlinear interferometers has drawn remarkable attention in recent years \cite{lemos2014quantum,kalashnikov2016infrared, zeilinger2017quantum,gilaberte2019perspectives,lindner2020fourier}. The transfer of these concepts to the terahertz frequency range is highly attractive, as the detection of terahertz radiation is challenging due to its inherently low photon energy. Even though the quantum-optical sensing and spectroscopy approach has been demonstrated in this spectral range \cite{kutas2020terahertz, kutas2021quantum}, its measurement time is not yet competitive. But the use of spontaneous parametric down-conversion and the correlation of the involved photons is not the only possibility to transfer photon properties to the visible range. Nonlinear upconversion has been proven to provide access to, e.g. the mid-infrared \cite{kehlet2015infrared,tidemand2016mid,israelsen2019real} or terahertz \cite{ding2006efficient,khan2007optical,khan2008optical,zaks2008single,pfeiffer2020terahertz} frequency range. We present a novel phase-sensitive frequency-conversion concept using nonlinear optics (up- and downconversion) in combination with interferometry of the generated photons with leftover photons of the used pulsed excitation light. Its potential is demonstrated by applying it to the measurement of sample thicknesses, where a standard deviation of 4~fs (corresponds to about 1.3~$\mu$ m on average) is achieved for measurement times below one second. The measurements are carried out for six different samples and are in good agreement with reference measurements performed using a standard time-domain spectroscopy (TDS) system \cite{weber2020influence}.

\section{Methods}
\subsection{Phase detection principle}
\begin{figure}[b!]
    \centering
    \includegraphics[width=\textwidth]{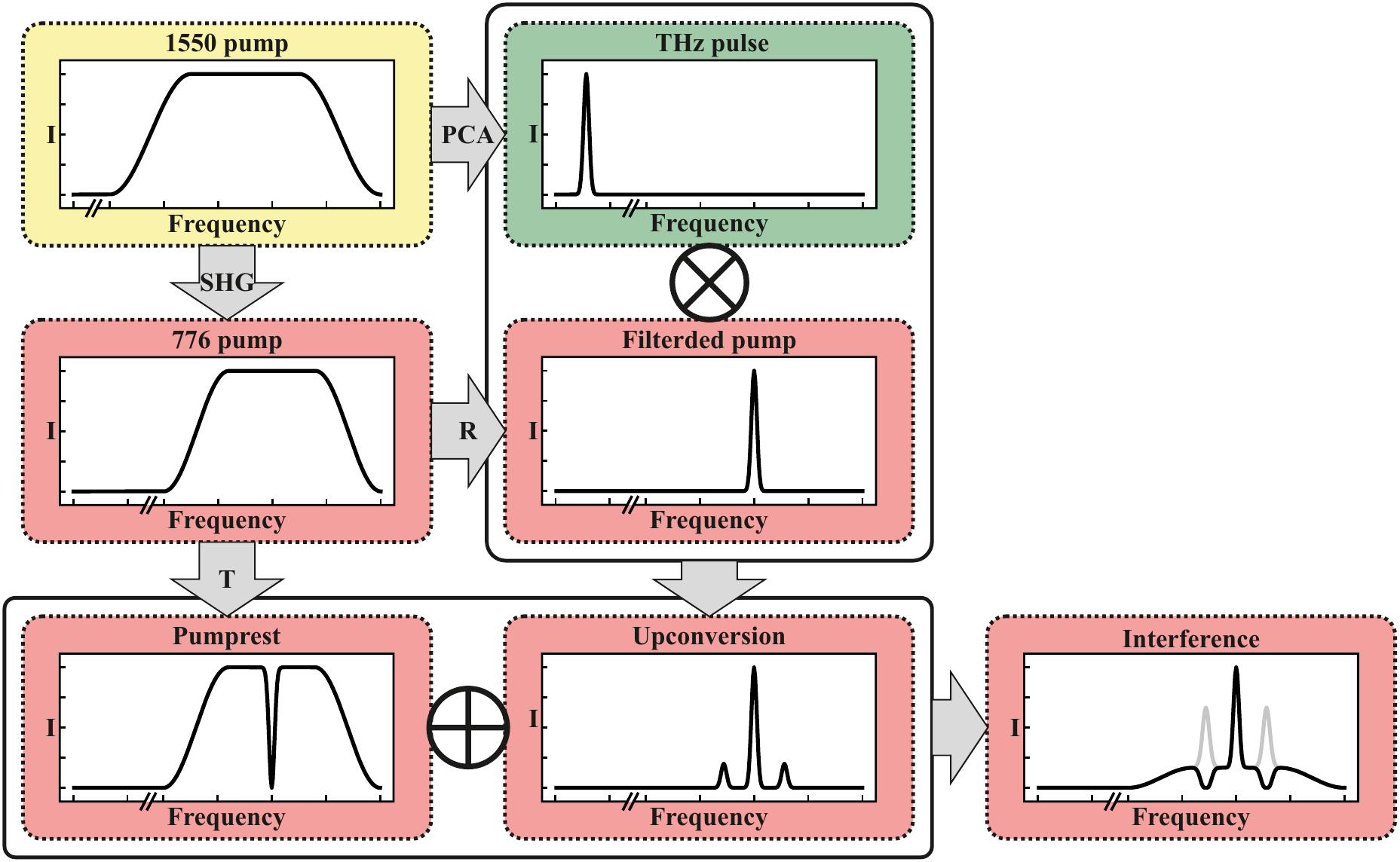}
    \caption{Scheme of the phase-sensitive terahertz upconversion detection: A single 1550~nm pump source is used to generate the terahertz radiation and a frequency-doubled pump radiation at 776~nm. This pump radiation is split by a filter into a narrow-bandwidth portion and a so-called pumprest (which was dumped in our earlier experiments). The narrow-bandwidth portion is mixed with the terahertz radiation, resulting in the generation of up- and downconversion sidebands. The superposition of these sidebands with the pumprest is the phase-sensitive interference, which is finally detected.}
    \label{Flowchart}
\end{figure}
Recently, we have demonstrated the intensity detection of terahertz radiation by frequency conversion into the visible range\cite{pfeiffer2020terahertz}. To detect not only the intensity but also the phase of the upconversion signal, a second source of light with identical wavelength and a fixed time and phase relation is superimposed with the upconverted signal to achieve interference. %The second light source must have a fixed time and phase relation to the upconverted signal. %While there are a lot of different approaches to obtain a source for interfering with the upconversion signal, 
Our proposed method is to use part of the pump light centered around 776~nm, as it spectrally extends into the upconversion region and inherits the required fixed phase and timing relation with respect to the pump photons. %this setup makes use of the share of pump light previously separated by the first RBP, further referenced as pumprest. Since the pumprest is separated from the optical pump pulse, it inherits the timing of the pump, while the separation allows control over the intensity of the pumprest using a gray wheel and the relative timing delay between pumprest and optical pump, using a piezo mounted mirror. 
To visualize the concept of our approach of phase-sensitive terahertz upconversion detection, Fig.~\ref{Flowchart} shows a diagram of all the spectral components involved in the experiment as well as their relation to each other. In the top left corner, the spectrum of the initial 1550~nm fs-pulsed (therefore spectrally broad) pump laser is sketched. The pulses are split to pump a photoconductive antenna (PCA) for terahertz generation and a second-harmonics generation (SHG) process to generate pump pulses at a center wavelength of about 776~nm. Using a reflective bandpass filter (RBP), a narrow spectral band of the 776~nm pump pulse is reflected from the RBP (R) while the remaining spectrum is transmitted (T). The reflected part, titled \emph{Filtered pump} in the diagram, is then superimposed with the terahertz radiation inside a nonlinear medium to generate the up- and downconversion signal. Combining the resulting conversion signal with the transmitted radiation from the RBP, titled \emph{Pumprest}, leads to interference between the two signals as shown in the bottom most node of the diagram. This interference signal is then phase-sensitive to the terahertz radiation.

%\section{Experimental setup}
\subsection{Setup}
\begin{figure}
\includegraphics[width = \textwidth]{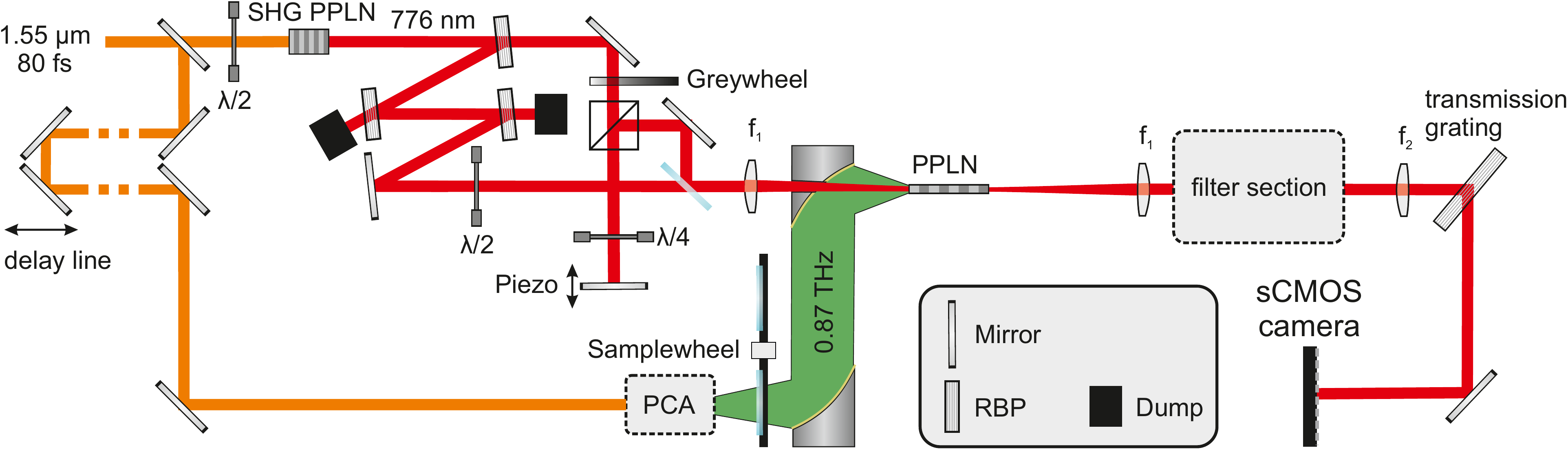}
\caption{Current measurement setup. A 1550~nm laser (orange) pumps the terahertz emitter and a frequency-doubling crystal to generate a NIR pump beam (red), which is spectrally narrowed down using reflective filters and superimposed with the terahertz radiation (green) before entering a nonlinear medium. The transmitted radiation of the first filter is merged with the optical pump with reduced intensity after reflecting off of a piezo controlled mirror. After the nonlinear medium, the pump wavelength is removed in a filter section and the remaining beam travels through a transmission grating onto a sCMOS sensor. SHG: second harmonics generation, PPLN: periodically poled lithium niobate, RBP: reflective bandpass, PCA: photoconductive antenna, $\mathrm{f_1, f_2}$: lenses (focal lengths of $\mathrm{f_1}$: 125~mm, $\mathrm{f_2}$: 400mm). \label{Experiment_Setup} }
\end{figure}

The setup used in this experiment is shown in Fig.~\ref{Experiment_Setup} and bases on the setup presented in \cite{pfeiffer2020terahertz}. It is running on a single erbium-doped fiber laser emitting femtosecond pulses in the 1550~nm range (orange) as pump source. This pump source is split in two branches, one of which is used to drive a second-harmonic generation (SHG) process in the first periodically poled lithium niobate (PPLN) crystal.
The frequency-doubled pulses emitted from the SHG crystal (red) are further spectrally narrowed down using reflective bandpass filters (RBP). In contrast to \cite{pfeiffer2020terahertz} and as described above, the share of the pump light which is transmitted by the first RBP, further referred to as pumprest, is not dumped but guided through a polarizing beam splitter (PBS) followed by a quarter-wave plate to be reflected on a piezo-controlled mirror. The reflected beam travels back through the quarter-wave plate into the PBS, which then reflects the beam due to the induced polarization rotation. A beam-sampling plate is used to recombine a fraction of the resulting beam with the spectrally filtered 776~nm pump pulse, which also overlaps temporally with the pumprest due to identical optical travel. The combined beam is then focused into a second PPLN, which is phase matched for SFG and DFG of the optical pump and terahertz radiation at 0.911~THz. The external terahertz radiation is supplied by a photoconductive antenna (PCA)\cite{dietz201364}, which is pumped by the second path of the original 1550~nm pump source. To overlap terahertz seed and optical pump pulse temporally, a delay line is incorporated in the second 1550~nm pump branch. The terahertz pulse emitted from the PCA is collimated and then focused into the PPLN using off-axis parabolic mirrors (OAPs), where the focusing mirror has a hole to allow the 776~nm pump to travel through, focusing on the same target. %It was found that combining terahertz seed and optical pump with a through-hole OAP reduces the losses for both components compared to the indium tin oxide (ITO) coated glass, which was used to combine the components in \cite{pfeiffer2020terahertz}, leading to an improved conversion efficiency inside the PPLN. 
After the crystal, the upconversion signal is separated from the remaining pump inside a filter section. %, which is done using a second set of RBPs. The filters are used in transmission to act as notch filters and remove the unconverted pump from the signal. 
Using a transmission grating, the spectrum of the conversion signal together with the pumprest are spectrally resolved and captured by an sCMOS sensor.

%\subsection{Piezo control parameters and measurement time}
%To cover a full period of the interference pattern, the total travel of the piezo for one sweep needs to be at least half of the pump wavelength $\mathrm{\lambda_P=776~nm}$, as the change in optical travel for the pumprest is double the piezo movement. In the current configuration, the piezo moves 13 steps, covering a distance of 481~nm and surpassing the requirement for a full interference period, which is indicated by the dashed red line. With a camera exposure time of 5~ms, the piezo reaches maximum travel after 65~ms and completes a full sweep in 130~ms, since both forward and backward travel are treated equally. 
\subsection{Data acquisition and evaluation}
\begin{figure}[t!]
\begin{center}
%\tikzsetnextfilename{CompareFigure}
\includegraphics[width=\textwidth]{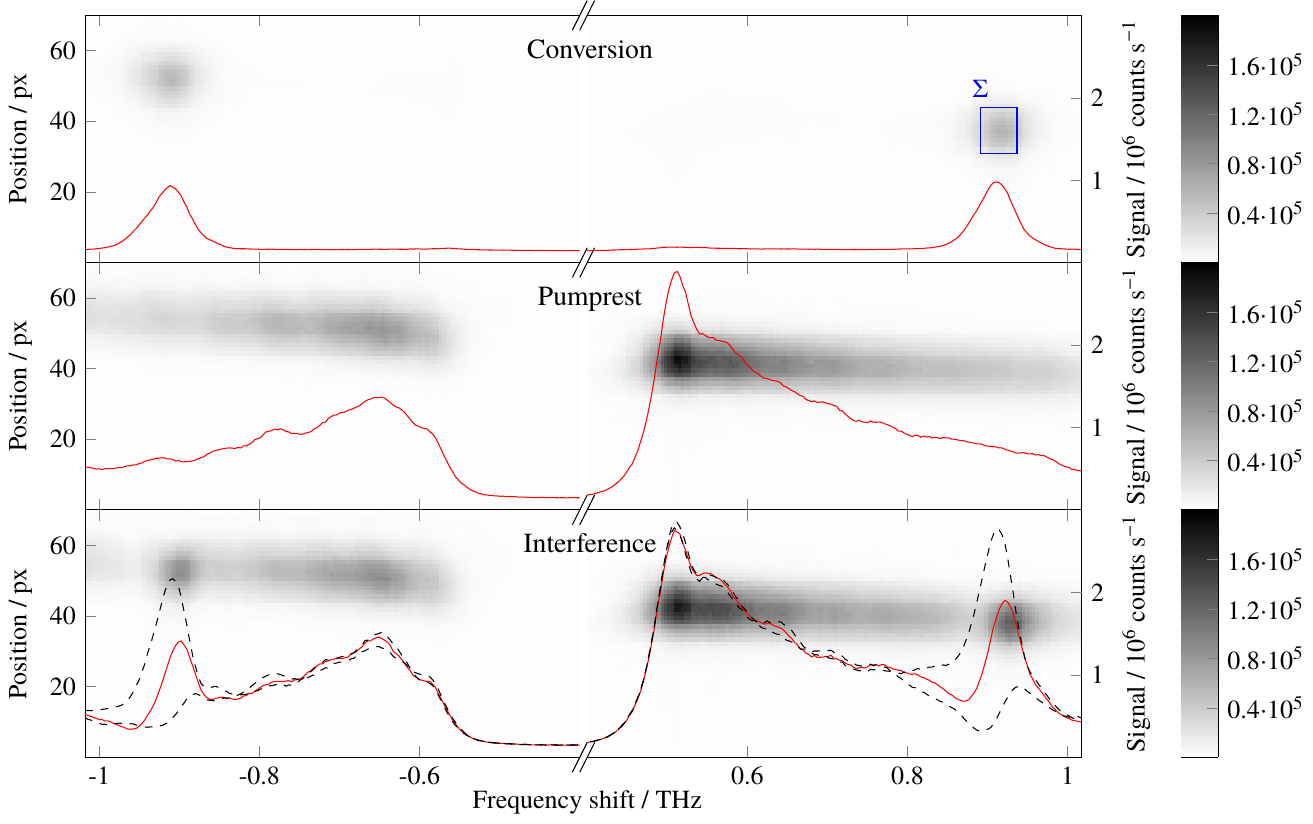}%{Tikz/Signal_compare/CompareFigure.tikz}
\end{center}
\caption{Camera images for the different signals. The top image shows the conversion signal only while the center image shows only the pumprest. in the bottom image, conversion and pumprest interfere. The red curve shows the column maximum of the image behind. The bottom image includes dashed curves for maximum constructive and destructive interference. \label{SignalCompare}}
\end{figure}

For a better understanding of the signal captured by the camera, the conversion signal and the pumprest are shown separately in Fig.~\ref{SignalCompare}. The top image shows the terahertz up- and downconversion signal as round Gaussian spots, with the downconversion on the left and the upconversion on the right side. The x-axis indicates the frequency shift of the signal in relation to the optical pump. %, which would appear in the center of the image without filtering. The filtering is necessary because the intensity of the remaining pump outweighs the intensity of the signal by multiple orders of magnitude, which would overexpose and potentially damage the image sensor. 
The red curve indicates the column sum for each image column, providing a cross section of the signal for a better visualization. In later steps, the integral of an area of 13x13 pixel surrounding the conversion maxima, indicated by the blue square, is used as measure of intensity, as done in \cite{pfeiffer2020terahertz}.
The camera signal of the pumprest with no conversion signal is shown in Fig.~\ref{SignalCompare} in the center image. %It is worth noting that the spectral intensity profile of the pumprest signal can not be tuned in the current setup, the symmetry of which is therefore not guaranteed. Although control over this profile could be gained using spatial light modulation in combination with gratings, the additional cost and complexity exceed the marginal benefit. 
The spectrum of the pumprest is much broader than the conversion signal, extending into the suppressed spectral range surrounding the pump. 
Overlapping the conversion signal with the pumprest results in interferometry between the two arms, as shown in the lower image in Fig.~\ref{SignalCompare}. While the red curve shows again the cross section of the camera image, the dashed black curves indicate the cross sections for optimal constructive and destructive interference, with the red curve in between. The impact of the phase-sensitive detection becomes apparent in the following section, where the signal is compared to the intensity-only measurement.
%The following section describes how the interference pattern is gathered and evaluated.%To evaluate the quality of the interference, these extrema are evaluated in the next section.

\subsection{Comparison to time-of-flight}
%The intensity of the conversion spots is dependent on the temporal overlap of the terahertz signal pulse and the optical pump pulse inside the PPLN. 
While the intensity of the upconversion signal alone can be used for time-of-flight measurements by varying the arrival time of the terahertz pulse as demonstrated in \cite{pfeiffer2020terahertz}, the difference in arrival time between the optical pump pulse and the terahertz pulse needs to be significant in order to detect the change in upconversion intensity. By measuring the phase of the upconverted signal as described, detection of smaller changes in optical path lengths is possible. Since the phase of the upconversion signal is directly correlated with the phase of the terahertz seed, all changes in the terahertz phase due to changes in the optical path length can be detected in the conversion signal. To compare the resolution of the phase-sensitive detection method with the intensity-only measurement from \cite{pfeiffer2020terahertz}, the delay between terahertz signal and optical pump is varied using the terahertz delay line, with the pumprest turned off to detect intensity only and turned on for the interferometric measurement. %In contrast to \cite{pfeiffer2020terahertz} where the delay line was driven in small steps to ensure maximum precision, for this measurement the delay line is moving at constant velocity while the camera captures signal data continuously. This is done to keep the overall measurement time low, which in turn reduces signal drift in the interference measurements.
\begin{figure}[b!]
\begin{center}
%\tikzsetnextfilename{IntensInter}
\includegraphics[width=\textwidth]{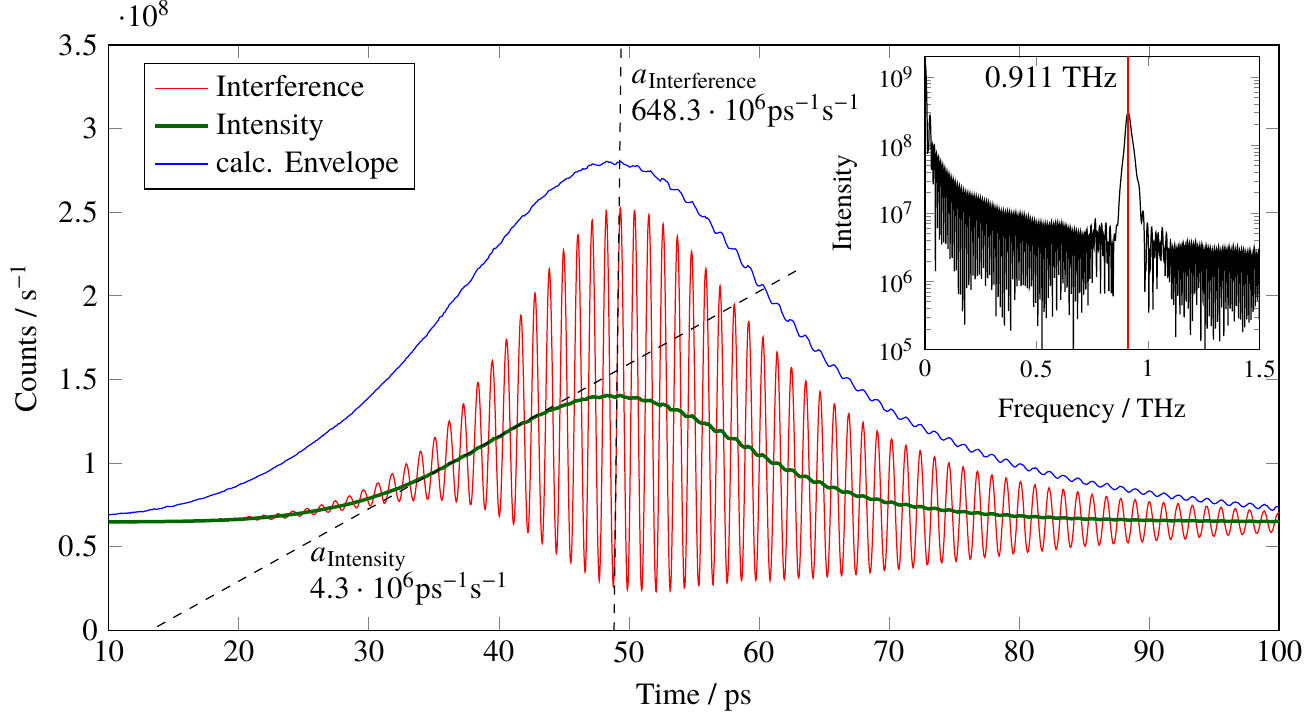}
\end{center}
\caption{Comparing the intensity-only signal with the phase-resolved signal when tuning the delay between optical pump and terahertz pulse. Due to the much steeper slopes on the interference curve, slight changes in position result in much larger changes in value. The blue curve indicates the calculated envelope of the interference curve. The interference measurement is also used to determine the frequency of the terahertz radiation.\label{IvsI}}
\end{figure}
In Fig.~\ref{IvsI}, the intensity-only signal is shown as green curve while the interference signal is displayed in red. For a better comparison of the signals, the intensity curve is offset by the constant pumprest signal. % It is worth noting that the theoretical envelope curve for the interference signal, shown in blue, is not only offset by the pumprest intensity but, according to Eq.~\eqref{eqmax}, has an additional term of $2 \sqrt{I_\mathrm{rest}} \sqrt{I_\mathrm{conv}}$. 
As a measure of resolution, the gradients for the steepest parts of the curves are used, as they express the maximum change in camera signal for a given delay in terahertz arrival. The gradients are also included in Fig.~\ref{IvsI} as dashed black lines. The calculated values for the gradients are $\mathrm{4.3\cdot10^6~ps^{-1}s^{-1}}$ for the intensity curve and $\mathrm{648.3\cdot10^6~ps^{-1}s^{-1}}$ for the interference curve. With this, the sensitivity of our proposed method with respect to optical path length changes is increased by a factor of about 150 in comparison to the intensity-only detection method. %While this measurement helps visualizing the impact of the additional phase information, the actual precision of the measurement is dependent on other aspects as well, which are discussed more deeply in the section below. 
Since the travel distance of the delay line used to measure the data from Fig.~\ref{IvsI} is well-known, the measurement can also be used to calculate the frequency of the interference signal, which equals the frequency of the terahertz signal, assuming that the phase of the pumprest is stable. %Good knowledge of the terahertz frequency is crucial in later steps, where the phase differences of the interference signal are used to determine the change in optical travel distance when a sample is inserted into the terahertz beam. 
The frequency is determined by Fourier transformation of the zero-filled interference signal and equals 0.911~THz. %In later steps, the interference pattern is measured by scanning the pumprest continuously using the piezo-mounted mirror, while the delay line for the timing of the terahertz signal is held constant. 

\subsection{Signal-to-pumprest ratio}
A standard parameter which characterizes the quality of interference is called visibility and describes the ratio between the theoretical and the measured difference between constructive and destructive interference. The visibility can reach at most 1.0 for a perfect interference setup. %To calculate the visibility, the interference maxima and minima have to be measured. A static phase between pumprest and upconversion signal results in a single reading of the interference signal as shown in Fig.~\ref{Visibility}~(a), with no information about the actual phase difference between both signals. To measure the dashed curves in the figure also, the phase of the pumprest is %The resulting interference signal is fitted with a sine function to acquire the signal extrema. 
%varied using the piezo-mounted mirror, until the integral of the image area reaches its maximum and minimum values, both of which are stored for evaluation.
Since theoretical extrema depend on the ratio between the isolated upconversion signal and the pumprest, both of these values are measured independently beforehand. The theoretical intensities of the interference extrema equal 
\begin{gather}
I_\mathrm{int,max} \propto (E_\mathrm{rest} + E_\mathrm{conv})^2 = E_\mathrm{rest}^2 +2 E_\mathrm{rest} E_\mathrm{conv} + E_\mathrm{conv}^2, \\
I_\mathrm{int,min} \propto (E_\mathrm{rest} - E_\mathrm{conv})^2 = E_\mathrm{rest}^2 -2 E_\mathrm{rest} E_\mathrm{conv} + E_\mathrm{conv}^2.
\end{gather}
Since the setup allows the measurement of intensities only, $E_\mathrm{rest}^2$ is replaced by $I_\mathrm{rest}$ and $E_\mathrm{conv}^2$ by $I_\mathrm{conv}$. This leads to the following equations:
\begin{gather}
I_\mathrm{int,max} = I_\mathrm{rest} +2 \sqrt{I_\mathrm{rest}} \sqrt{I_\mathrm{conv}} + I_\mathrm{conv}, \label{eqmax} \\
I_\mathrm{int,min} = I_\mathrm{rest} -2 \sqrt{I_\mathrm{rest}} \sqrt{I_\mathrm{conv}} + I_\mathrm{conv}. \label{eqmin}
\end{gather}

\begin{figure}
\begin{center}
%\tikzsetnextfilename{DescrFigure}
\includegraphics[width=\textwidth]{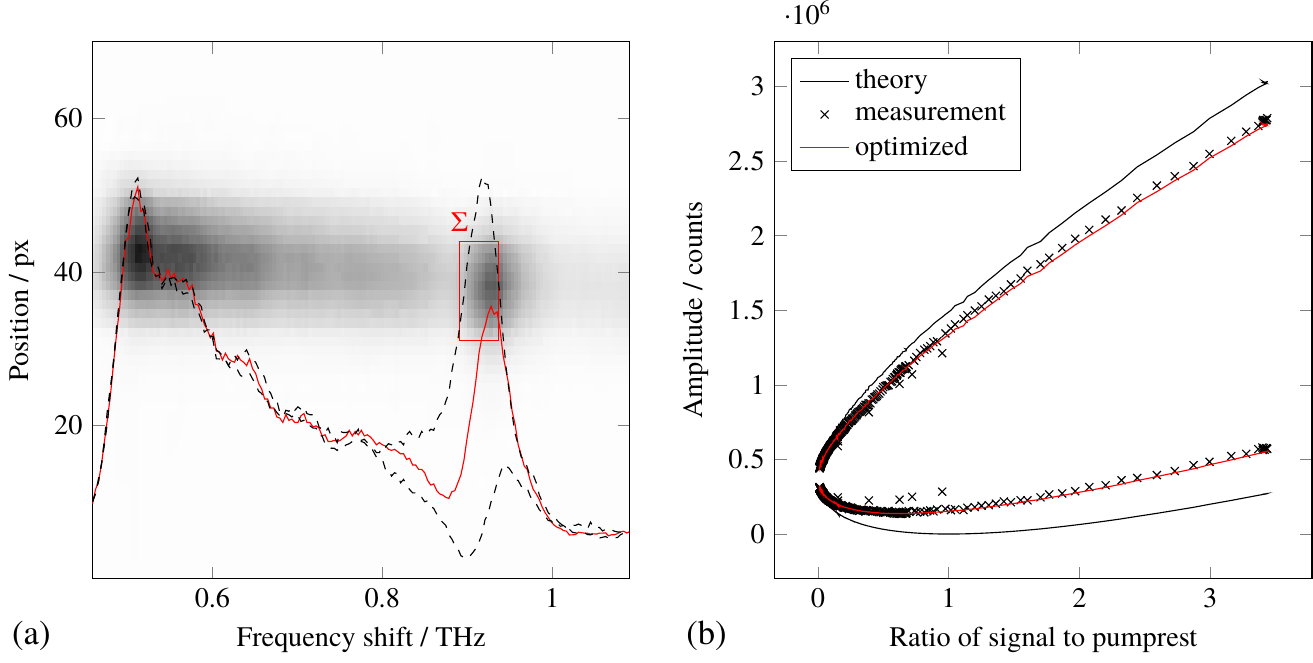}
\end{center}
\caption{Evaluation of the interference visibility. The diagram (a) shows a cutout of Fig.~\ref{SignalCompare} including a rectangle indicating the integrated image area for evaluation. Diagram (b) shows minimum and maximum integrals for a given ratio of conversion signal to pumprest as black marks with the theoretical values as black curves. The red curves show the theoretical values including an additional correction factor.\label{Visibility}}
\end{figure}

Using this formula, the theoretical extrema are calculated for different ratios and displayed as black lines in Fig.~\ref{Visibility}~(b). For a ratio of 1.0, the destructive interference curve reaches its minimum at zero intensity, which indicates perfect destructive interference. The difference between the maximum and minimum curve increases with a higher ratio, following the equation
\begin{equation}
I_\mathrm{\Delta int} = 4 \sqrt{I_\mathrm{rest}} \sqrt{I_\mathrm{conv}}.
\end{equation}
The black marks in Fig.~\ref{Visibility}~(b) show the measured minima and maxima for the given ratios. While the data points form a similar shape, the measurement does not align with the theory. Also, the minimum curve does not reach its lowest point at a ratio of 1.0, which indicates that the calculated ratios do not match the actual ratio of interfering radiation.
A possible source for this mismatch are the temporal profiles of the interfering pulses. By using the camera signal to measure the intensity of each component, only the average intensity can be determined. If the pulse of one of the components is shorter in time but has a higher peak intensity, the camera signal will be identical for the individual components, the interference however will not be ideal because part of the longer pulse has no temporal overlap with the shorter pulse. Other effects such as spatial overlap or a chirped pulse might result in similar deviations. To account for this effect, an additional factor $a$ is introduced to equations \eqref{eqmax} and \eqref{eqmin} as follows
\begin{gather}
I_\mathrm{int,max} = I_\mathrm{rest} +2 a \sqrt{I_\mathrm{rest}} \sqrt{I_\mathrm{conv}} + I_\mathrm{conv},\label{eqmaxcorr} \\
I_\mathrm{int,min} = I_\mathrm{rest} -2 a \sqrt{I_\mathrm{rest}} \sqrt{I_\mathrm{conv}} + I_\mathrm{conv},\label{eqmincorr}
\end{gather}
which describes the share of one component which can interfere with the other. The red curve in Fig.~\ref{Visibility} shows the calculated curve using a parameter of $a=0.795$, which minimizes the accumulated deltas between the measured values and the calculated curve and matches the measurement to a high degree. The additional parameter $a$ corresponds to the visibility which, with a value of 0.795, provides sufficient contrast for the desired use case.

%consecutive measurements of the additional time of flight of inserted samples with a precision of <4~fs.

%%%%%%%%%%%%%%%%% copypasta von unten
%\subsection{Phase determination}
\section{Results}
\subsection{Sample Measurements}

\begin{figure}
%\begin{center}
%\tikzsetnextfilename{PhaseDescr}
\includegraphics[width=\textwidth/2]{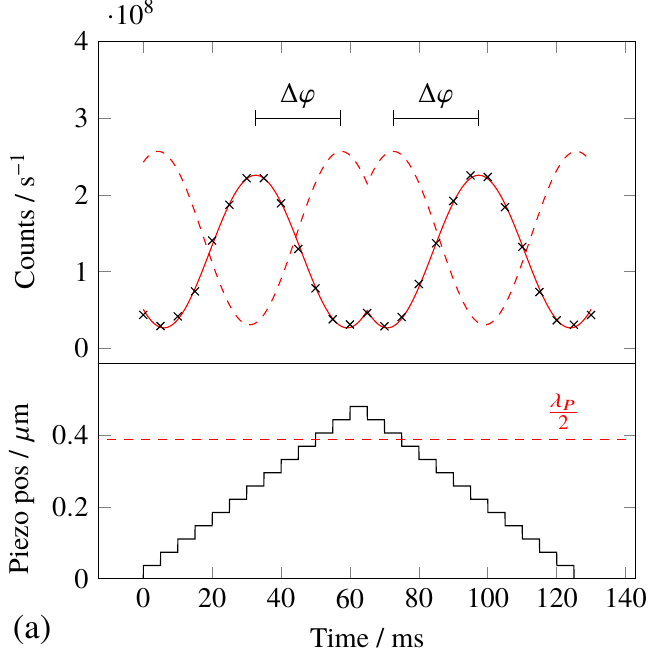}
%\tikzsetnextfilename{Longterm}
\includegraphics[width=\textwidth/2]{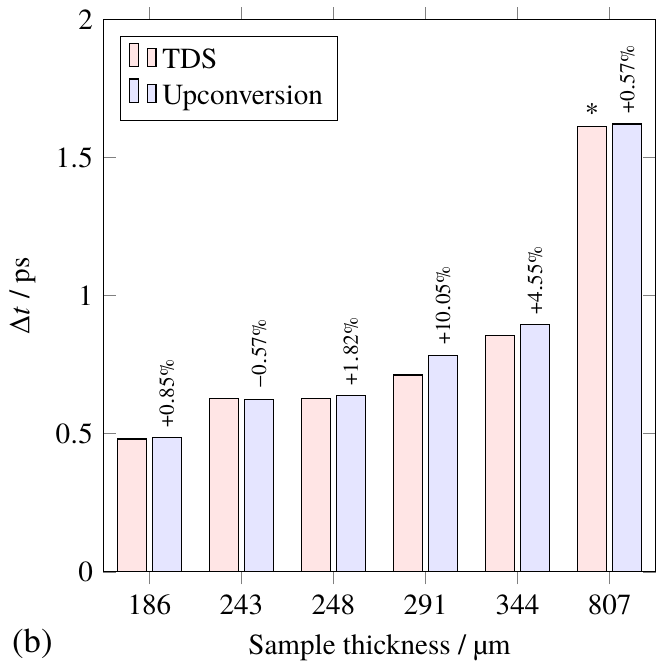}
%\end{center}
\caption{(a) Diagram of the phase measurement process. The piezo moves forward and backward while the upconversion intensity is measured. %During sample insertion, the measurement is ignored. (b) Long-term measurement of the phase of upconversion (red) and downconversion signal (blue), displayed together with the average of both signals in black. The inset shows a magnified part of the signal, where a symmetric short-term behavior of the phases can be perceived. 
(b) Time-of-flight measurement results for PET-samples with different thicknesses compared to TDS measurements. For the last sample, the phase shift exceeded $2\pi$ due to the high thickness, which was compensated accordingly afterwards.\label{Results}}
\end{figure}

By periodically moving the piezo-mounted mirror, the phase of the pumprest in relation to the conversion signal is varied constantly. For linear movement of the piezo, this results in a sine-wave-like interference pattern, which can be sampled using the camera. In Fig.~\ref{Results}~(a), the stepwise movement of the piezo is shown in the lower part of the diagram, while the data acquired from the camera (pixel intensity integral of the area of interference) is displayed in the upper part. %To prevent movement of the piezo during camera exposure and assure linear movement, the piezo does not move continuously but in steps of 37~nm, each of which is hardware triggered by a signal from the camera indicating the end of exposure. It is worth noting that although the piezo operates in an open loop configuration, the steps are considered equidistant, ignoring possible nonlinear positioning of the piezo. A position control using strain gauge readout was implemented but the uncertainty of the readout due to the short integration time introduced a higher margin of error than using the applied voltage to calculate the position. 
%To cover a full period of the interference pattern, the total travel of the piezo for one sweep needs to be at least half of the pump wavelength $\mathrm{\lambda_P=776~nm}$, as the change in optical travel for the pumprest is double the piezo movement. In the current configuration, the piezo moves 13 steps, covering a distance of 481~nm and surpassing the requirement for a full interference period, which is indicated by the dashed red line. With a camera exposure time of 5~ms, the piezo reaches maximum travel after 65~ms and completes a full sweep in 130~ms, since both forward and backward travel are treated equally. 
A sine function is fitted to the interference pattern (shown as red curve), whose phase is used to detect changes in the terahertz beam path, as it correlates with the phase of the upconversion signal and therefore the terahertz phase.  %To reduce computational effort, the wavelength of the sine fit is evaluated only in the beginning of the measurement in a separate calibration process and is held constant for later evaluations. 
%The phase of the sine fit can now be used to detect changes in the terahertz beam path, as it correlates with the phase of the upconversion signal and therefore the terahertz phase. 
%Although the absolute value of the evaluated phase is arbitrary, it remains constant unless external influences change the timing (phase) relations between the interfering components. A closer discussion of those influences can be found in the methods section. 
When inserting a sample into the terahertz beam, the phase of the interference pattern shifts. The dashed red curve shows the sine fit of the pattern before the sample is inserted, which is shifted by $\Delta \varphi$ relative to the solid curve.
%\subsection{Sample measurements}
Using an automated sample wheel, an array of six samples of PET plastic with precisely determined physical thicknesses ranging from 186~$\mu$ m to 807~$\mu$ m can be inserted into and removed from the terahertz beam to measure the phase shift.
A single measurement consists of four piezo sweeps, where the reference and sample measurements take a single sweep each and the insertion process takes two sweeps in between. The total measurement time is therefore specified as 520~ms. %It is worth noting that both forward and backward movements of the piezo are evaluated, which does not lead to a reduction in measurement time but results in two separate measurement values for a single measurement cycle as described. 
For statistical purposes, the sample wheel completes 100 full rotations, inserting and removing each of the six samples every rotation. 
%\begin{figure}
%\includegraphics[width=\textwidth/2]{Externalize/WheelImg.pdf}
%\includegraphics[width=\textwidth/2]{Externalize/MeasData.pdf}
%
%\caption{Image of the sample wheel (a) and measurement results (b). The sample wheel is used to insert samples into the terahertz beam, which changes the phase of the interference signal. The phase change is divided by the terahertz wavelength to acquire the additional time  of flight for the terahertz radiation, which is then compared to TDS measurements in (b). Due to the thickness, a factor of $2\mathrm{\pi}$ is added to the measured phase for $\ast$.\label{Results}}
%\end{figure}
With the wavelength of the terahertz radiation, the averaged $\Delta \varphi$ values for each sample can be used to calculate the additional optical path length of the terahertz radiation introduced by the sample. These values are compared to reference measurements performed with a state of the art terahertz TDS measurement system\cite{weber2020influence}. Figure~\ref{Results} shows the results for all samples as a bar plot, with the reference TDS measurement results in blue and the results of the upconversion measurement in red.

\subsection{System stability}

\begin{figure}
    \centering
    \includegraphics[width=\textwidth]{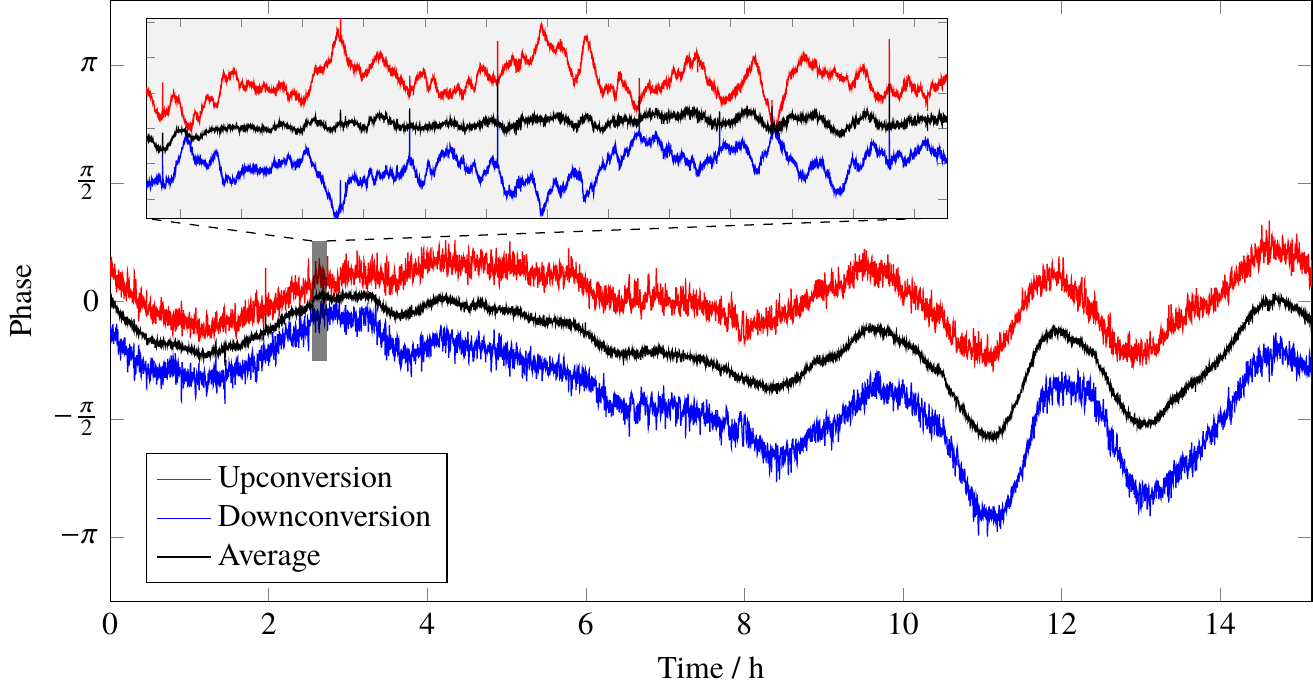}
    \caption{Long-term measurement of the phase of upconversion (red) and downconversion signal (blue), displayed together with the average of both signals in black. The inset shows a magnified part of the signal, where a symmetric short-term behavior of the phases can be perceived. }
    \label{Longterm}
\end{figure}

To measure the stability of the interference phase over time, the phase of both up- and downconversion was continuously evaluated over a 15~h period. The results are shown in Fig.~\ref{Longterm}, where the red curve indicates the phase of the upconversion signal and the blue curve represents the downconversion. The black curve shows the average value between up- and downconversion. Considering the full 15~h measurement window, the red and blue curve seem to drift mostly in equal direction, moving the average curve with them. This behavior is assumed to be caused by thermal influence on the path of the pumprest and is identical to the effect of moving the piezo mounted mirror. This effect can be compensated for by using the average signal as artificial zero line.
The inset in Fig.~\ref{Longterm} however shows additional symmetrical drifts which do not affect the average curve. These drifts are caused directly by variations of the terahertz pulse timing in relation to the optical pump and also originate from thermal influences, which act on the optical fiber paths from the shared pump to the emitters of the terahertz and pump radiation. These drifts affect the interference pattern in the same fashion as a real measurement would. When measuring samples however, the samples can be inserted and removed from the terahertz beam quickly enough to neglect the influence of the drift.
%\newpage

%\subsection{Possible sources of errors}
%A possible source of systematic error would be that the sample is not inserted in the collimated area of the terahertz beam but in the diverging beam immediately after the emitter. This leads to slightly longer optical travel for parts of the radiation and could lead to an overestimation of the actual travel path. While almost all of the measurements are slightly overestimated, the error would be a constant percentage for all samples. The perceived error also does not seem to be connected to the material thickness, as the sample with the highest thickness shows the least deviation of all evaluated samples. This leads to sample-specific properties which influence the evaluation. High dispersion of the material in the terahertz range could lead to different results when different terahertz spectra are used, although the spectrum of the TDS system in use has its spectral maximum at around 1~THz as well.

\section{Discussion}
The determined thicknesses using the novel phase-sensitive upconversion method agree well with the results of the TDS measurement system. While the results of two samples show greater deviations, the other samples deviate only 1\% on average. %While samples 1-3 with thicknesses between 186~\textmu m and 248~\textmu m and sample 6 with a thickness of 807~\textmu m show very narrow differences between the measurement and the reference value, sample 5 (344~\textmu m) and sample 4 (291~\textmu m) especially deviate substantially from the reference. 
Due to the limited aperture of the samples, the samples are inserted into the diverging part of the terahertz beam, which leads to a slight increase in optical travel through the sample and therefore a systematic error. Although the evaluated thicknesses show the tendency to be overestimated, a correlation between sample thickness and the deviation cannot be seen. The two outliers must therefore be connected to sample-specific properties which influence the evaluation. High dispersion of the material in the terahertz range could lead to different results when different terahertz spectra are used, although the spectrum of the TDS system in use has its spectral maximum at around 1~THz as well.
It is worth noting that sample 6 causes the phase to shift by more than $2\pi$ due to the high thickness of the sample. The measured phase shift was compensated accordingly. The standard deviation of 100 measurements for each sample is below 4~fs, which corresponds to a an average deviation of 1.3~$\mu$ m in terms of thickness. Since the first sample is the thinnest at 186~$\mu$ m, the precision of the measurement has the most impact on the result, causing a relative uncertainty of 0.6~\%. Due to the small deviations, Fig.~\ref{Results}~(b) does not include error bars. %A possible source of systematic error would be that the sample is not inserted in the collimated area of the terahertz beam but in the diverging beam immediately after the emitter. This leads to slightly longer optical travel for parts of the radiation and could lead to an overestimation of the actual travel path. While almost all of the measurements are slightly overestimated, the error would be a constant percentage for all samples. The perceived error also does not seem to be connected to the material thickness, as the sample with the highest thickness shows the least deviation of all evaluated samples. This leads to sample-specific properties which influence the evaluation. High dispersion of the material in the terahertz range could lead to different results when different terahertz spectra are used, although the spectrum of the TDS system in use has its spectral maximum at around 1~THz as well. 

So far, the proposed method is realized for a narrow spectral bandwidth of the detected terahertz photons (due to the used phase-matching scheme), but already proves the applicability of the proposed method. By implementing other nonlinear media or phase-matching schemes, the spectral bandwidth can be enhanced, which might enable the measurement of multilayer structures in reflection (which is one of the demands of many applications). 

\section{Conclusions}
To conclude, we have introduced a novel phase-sensitive frequency-conversion detection principle that enables fast measurements in the terahertz frequency range by only detecting visible light. By implementing an interferometry principle, the sensitivity for optical path length changes is increased by a factor of 150 in comparison to intensity-only frequency-conversion methods. The detected interference provides a good visibility of 0.8 and the high short-term stability of the interference pattern allows for thickness measurements within less than a second. Measurements on PET samples demonstrated a standard deviation lower than 4~fs, corresponding to below 0.6 \% in thickness using measurement times below one second. Future extensions to two-dimensional measurements will elevate the potential of this approach.  
%In this work we demonstrated the extension of our previous setup to achieve phase-sensitive detection of terahertz-radiation using upconversion. With this, the sensitivity for optical path length changes is increased by a factor of 150. The detected interference provides a good visibility of 0.8 and the short-term stability of the interference pattern allows for thickness measurements. Measurements on PET samples demonstrated a standard deviation lower than 4~fs, corresponding to below 0.6 \% in thickness using measurement times below one second. Future extensions to two-dimensional measurements will elevate the potential of this approach.  

%%%%%%%%%%%%%%%%%%% ende copypasta

\section*{Funding}
This work was carried out in the scope of the Fraunhofer Lighthouse Project "Quantum Methods for Advanced Imaging Solutions" (QUILT).

\section*{Disclosures}
The authors declare no conflicts of interest.

\section*{Data availability}
The data that support the findings of this study are available from the corresponding 
author upon reasonable request.

%\section*{Author contributions}
%G.v.F. and D.M. initiated this research. T.P. and D.M. designed the experiment. T.P. carried out the experiment. J.K. provided the reference data.

\end{document}